\documentclass[12pt]{article}

\usepackage{amssymb}
\usepackage{amsthm}
\usepackage{amsmath}
\numberwithin{equation}{section}
\usepackage{graphicx}

\newtheorem{theorem}{Theorem}[section]

\newtheorem{remark}[theorem]{Remark}

\newcommand{\D}{\mathrm{d}}

\newcommand{\uu}{{\underline{u}}}
\newcommand{\pp}[2]{ \frac{\partial #1}{\partial #2} }
\begin{document}
\title{On universality of homogeneous Euler equation}
\author{
B.G.Konopelchenko $^1$ and G.Ortenzi $^{2}$ 
\footnote{Corresponding author. E-mail: giovanni.ortenzi@unimib.it,  Phone: +39(0)264485725 }\\
$^1$ {\footnotesize Dipartimento di Matematica e Fisica ``Ennio De Giorgi'', Universit\`{a} del Salento, 73100 Lecce, Italy} \\
 $^2$ {\footnotesize  Dipartimento di Matematica e Applicazioni, 
Universit\`{a} di Milano-Bicocca, via Cozzi 55, 20126 Milano, Italy}\\
$^2${\footnotesize  INFN, Sezione di Milano-Bicocca, Piazza della Scienza 3, 20126 Milano, Italy}
} 
\maketitle
\abstract{Master character of the multidimensional homogeneous Euler equation is discussed. It is shown that under restrictions to the lower dimensions certain subclasses of its solutions provide us with the solutions of various hydrodynamic type equations. Integrable one dimensional systems in terms of Riemann invariants and its extensions,
 multidimensional equations describing isoenthalpic and polytropic motions and shallow water type equations are among them. }
\section{Introduction}
Homogeneous Euler equation (also called pressureless Euler  equation)
\begin{equation}
\pp{u_i}{t}+\sum_{k=1}^n u_k \pp{u_i}{x_k}=0\, , \qquad i=1,\dots,n\, ,
\label{plessE}
\end{equation}
is one of the basic equations in the theories of fluids, gas and other media at $n=3$ (see e.g. \cite{L-VI, Lamb,Whi}). In spite of the fact that it represents the most simplified version (no pressure, no viscosity etc. \cite{L-VI,Lamb}) of the full equations, it arises in number of studies in many branches of physics. \par

Euler equation (\ref{plessE}) has the remarkable property to be solvable by the straightforward multi-dimensional extension of the classical hodograph
equations method \cite{Fai1,Fai2}. This fact and reductions to lower number of dependent variables has been used to establish the interrelations between equations 
(\ref{plessE}) and multi-dimensional Monge-Amp\`ere equations and Bateman equations \cite{Fai2,FL1,FL2,Lez}. \par

In the present paper we will study the restrictions of the $n$-dimensional Euler type equation to the lower dimensional spaces  $\mathbb{R}^m$ ($m < n$). We start
with the slightly modified equation (\ref{plessE}), namely, with the system
\begin{equation}
 \pp{u_i}{t}+\sum_{k=1}^n \frac{\beta_k}{\alpha_k} \lambda_k(\uu) \pp{u_i}{x_k}=0\, , \qquad i=1,\dots,n\, ,
\label{plessE-g}
\end{equation}  
where $\lambda_k(\uu)=\lambda_k(u_1,\dots,u_n) $ are arbitrary real-valued functions and $\alpha_k,\beta_k$ are arbitrary real constants. Solutions of the system 
(\ref{plessE-g}) are provided by the hodograph equations
\begin{equation}
\alpha_i x_i -\beta_i \lambda_i(\uu) t+ f_i(\uu)=0\, , \qquad i=1,\dots,n\, ,
\label{hodo-g-intro}
\end{equation}
where  $f_k(\uu)=f_k(u_1,\dots,u_n) $ are arbitrary real-valued functions (in the case $\lambda_i(\uu)=u_i$ see \cite{Fai1,Fai2}). Constants $\alpha_i,\beta_i$ are, obviously,
transformable away except the cases when some of them vanish.  Exactly such cases are related with  restrictions of the system (\ref{plessE-g}).\par

It is shown  that restrictions on independent variables $x_i$, functions $f_i$ and parameters $\alpha_i,\beta_i$ in (\ref{hodo-g-intro}) gives rise to the various hydrodynamical
type systems in the spaces $\mathbb{R}^m$ ($m \leq n$). In particular, under the restriction $x_1=x_2=\dots=x_n\equiv x$ plus certain restrictions on $g_i$, 
the hodograph equations (\ref{hodo-g-intro}) provides us with the solutions of the system
\begin{equation}
\pp{u_k}{t}+\lambda_i(\uu) \pp{u_i}{x}=0\, , \qquad i=1,\dots , n\, ,
\label{cons-R}
\end{equation} 
that is the classical diagonalized one-dimensional system in terms of Riemann invariants solvable by Tsarev's generalized hodograph method  \cite{Tsa,DN}.\par 

Under the restriction to the $(n-1)$-dimensional subspace given by $x_n=0$, with constraints on functions $f_i$, the hodograph equations (\ref{hodo-g-intro}) provide us 
with solutions of $(n-1)$-dimensional hydrodynamical type systems of certain interest. First example is given by equations
\begin{equation}
\begin{split}
&\pp{u_i}{t}+\sum_{k=1}^{n-1} u_k \pp{u_i}{x_k}=-\pp{v}{x_i}\, , \qquad i=1,\dots,n-1\, , \\
&\pp{v}{t}+\sum_{k=1}^{n-1} u_k \pp{v}{x_k}=0\, ,
\end{split}
\label{nD-Jor}
\end{equation}
which describes the adiabatic and isoenthalpic motion where $v=TS$ and $T$ is temperature and $S$ is entropy. 
In the particular case $f_i=\pp{W}{u_i}$, $i=1,\dots,n$, this system describes the potential motion $\left( u_i=\pp{\phi}{x_i} \right)$.
Second case describes the polytropic motion, namely,
\begin{equation}
\begin{split}
&\pp{u_i}{t}+\sum_{k=1}^{n-1} u_k \pp{u_i}{x_k}=-\frac{1}{\rho}\pp{p}{x_i}\, , \qquad i=1,\dots,n-1\, , \\
&\pp{\rho}{t}+\sum_{k=1}^{n-1}  \pp{}{x_k}(\rho u_k)=0\, ,
\end{split}
\label{nD-poly}
\end{equation}
where the density $\rho=u_n$ and the pressure $p=\rho^\gamma$. In this case the functions $f_i=\pp{W}{u_i}$ $i=1,\dots,n$ 
and the function $W$ obeys a determinant type PDE.\par

Natural two and higher dimensional extensions of the system (\ref{cons-R}) are considered too. \par

It is noted that solutions of the systems (\ref{cons-R})-(\ref{nD-poly}) and others are obtained in our approach are given by subclasses of solutions 
of the original system (\ref{plessE-g}) which are characterized by the specific choices of functions $f_i$ and restrictions on the coordinates $x_1, \dots, x_n$. \par

Interrelations between infinite-dimensional Euler equation (\ref{plessE-g}) and Burgers and Korteweg-de Vries equations is considered. 
We have also  discussed the phenomenon of the gradient catastrophe for homogeneous Euler equation. It is shown that it 
first happens at a point on the $n$-dimensional hypersurface.\par

The paper is organized as follows. In section \ref{sec-hodoE}  hodograph equations for the generalized homogeneous Euler equations and its one-dimensional 
reductions are considered. The $(n-1)$-dimensional reductions of the homogeneous Euler equations to the $(n-1)$-dimensional equations describing isoenthalpic motion and  
its potential version are studied in sections \ref{sec-Jordan} and \ref{sec-potJ}.  Reduction describing polytropic motion is discussed \ref{sec-Poly}. 
Reductions of the three-dimensional Euler equation are considered in section 
\ref{sec-3Ered}. Interrelation between infinite-dimensional Euler equation and Burgers and Korteveg-de Vries equations is analyzed in section \ref{sec-infEred}.
In section \ref{sec-catE} we discuss the gradient catastrophe for the homogeneous Euler equation.
\section{Generalized homogeneous Euler equation and its one-dimensional reductions}
\label{sec-hodoE}
We start with hodograph equations 
\begin{equation}
\alpha_i x_i -\beta_i \lambda_i(\uu) t+ f_i(\uu)=0\, , \qquad i=1,\dots,n\, ,
\label{hodo-g}
\end{equation}
where $\lambda_i(\uu)$ and $f_i(\uu)$ are arbitrary real valued  functions and $\alpha_,\beta_i$ are arbitrary constants. The system (\ref{hodo-g}) 
is an obvious extension of the hodograph equations considered in \cite{Fai1, Fai2, FL1}. \par

Differentiating (\ref{hodo-g}) w.r.t. $x_k$ and $t$, one obtains
\begin{equation}
\alpha_i \delta_{ik}+\sum_{l=1}^n\pp{g_i}{u_l}\pp{u_l}{x_k}=0\, , \qquad i,k=1,\dots, n .
\label{hododx}
\end{equation}
and
\begin{equation}
-\beta_i \lambda_{i}+\sum_{l=1}^n\pp{g_i}{u_l}\pp{u_l}{t}=0\, , \qquad i,k=1,\dots, n . 
\label{hododt}
\end{equation}
where $g_i \equiv \alpha_i x_i -\beta_i \lambda_i(\uu) t +f_i(\uu)$. \par

Relations (\ref{hododx}) and (\ref{hododt}) imply that
\begin{equation}
\pp{u_l}{x_k}=-(A^{-1})_{lk} \alpha_k \, , \qquad i,k=1,\dots, n .
\label{udx}
\end{equation}
and
\begin{equation}
\pp{u_l}{t}= \sum_{k=1}^n (A^{-1})_{lk} \beta_k \lambda_k\, , \qquad l=1,\dots, n .
\label{udt}
\end{equation}
where the matrix $A$ has elements
\begin{equation}
A_{lk}=\pp{g_l}{u_k}=-t \pp{\lambda_l}{u_k}+\pp{f_l}{u_k}\, , \qquad l.k=1,\dots,n 
\label{A-diff}
\end{equation}
and it is assumed that $\det(A) \neq 0$. \par

Combining (\ref{udx}) and (\ref{hododx}), one gets
\begin{equation}
\pp{u_l}{t}+\sum_{k=1}^n \frac{\beta_k}{\alpha_k} \lambda_k \pp{u_l}{x_k}=0\, , \qquad l=1,\dots,n\, .
\label{g-Hopf-g}
\end{equation}
For $\lambda_k=u_k$ , $\alpha_k=\beta_k=1$ these calculations has been done in \cite{Fai1, Fai2, FL1}.\par

Relations (\ref{hododx}) and (\ref{hododt}) also imply that 
\begin{equation}
\sum_{l=1}^n  A_{il} \left(\pp{u_l}{t}+\sum_{k=1}^n \frac{\beta_k}{\alpha_k} \pp{u_l}{x_k}\right) = 0\, , \qquad i=1,\dots,n\, .
\label{A-Hopf-g}
\end{equation}
It is noted that functions $f_i(\uu)$ are arbitrary one in this construction. They are related to initial data at $t=0$ {\it via}
\begin{equation}
\alpha_i x_i + f_i(\uu(t=0))=0\, . 
\label{0hodo}
\end{equation}
Hence, hodograph equations (\ref{hodo-g}) provide us with the general solutions of the system (\ref{g-Hopf-g}).\par

We would like to note that the system (\ref{g-Hopf-g}) and the original Euler equation (\ref{plessE}), in fact, are equivalent. Indeed it is easy to see that if $u_i$
obey the system (\ref{g-Hopf-g}) then $\lambda_k$ obey the system (\ref{plessE}) and viceversa ($u_i \to \lambda_i(\uu)$). It is noted also that the systems (\ref{g-Hopf-g})
with  different $\lambda_k(\uu)$ pairwise commute. So one has an infinite hierarchy of equations of the form (\ref{g-Hopf-g}).\par 

Now let us consider the simplest reductions of the system (\ref{g-Hopf-g}) for which the matrix $A$ is diagonal one, i.e.
\begin{equation}
A_{lk} = -t \pp{\lambda_l}{u_k}+ \pp{f_l}{u_k}=0 \, , \qquad l \neq k\, .
\label{A-cond}
\end{equation}  
 A way to satisfy this condition is to impose the constraints
 \begin{equation}
 \pp{\lambda_l}{u_k}=0\, , \qquad \pp{f_l}{u_k}=0\, , \qquad l \neq k\, .
 \end{equation}
 In this case the relations (\ref{hododx}) imply  that
 \begin{equation}
 \pp{u_l}{u_k}=0 \, , \qquad l \neq k\, ,
 \end{equation}
 and the $n$-dimensional system (\ref{g-Hopf-g}) (with $\alpha_k=\beta_k=1$) is decomposed into $n$ decoupled one-dimensional Burgers-Hopf type equations
 \begin{equation}
 \pp{u_l}{t}+\lambda_l(u_l) \pp{u_l}{x_l}=0\, , \qquad l=1,\dots,n\, .
 \end{equation}
 
 Less trivial reduction with the diagonal matrix $A$ arises if one considers the restriction to the one-dimensional subspace given by the condition 
 $x_1=x_2=\dots=x_n \equiv x$. In this reduction, the hodograph system (\ref{hodo-g}) assumes  the form 
 \begin{equation}
 x-\lambda_i(\uu)\, t+f_i(\uu)=0\, , \qquad  i=1,\dots,n\,  ,
 \label{hodo1D}
 \end{equation}
 and one obtains
 \begin{equation}
 \pp{u_l}{x}=-\frac{1}{\pp{f_l}{u_l}}\, , \qquad \pp{u_l}{t}=\frac{\lambda_l}{\pp{f_l}{u_l}}\, ,  \qquad l=1,\dots,n\,  .
 \end{equation}
 Hence, one has the system
 \begin{equation}
 \pp{u_l}{t}+\lambda_l(\uu) \pp{u_l}{x}=0\, , \qquad l=1,\dots,n\, .
 \label{Rinv1D}
 \end{equation}
 The equation (\ref{hodo1D}) implies that 
 \begin{equation}
 t=\frac{f_i-f_l}{\lambda_i-\lambda_l}\, ,\qquad i \neq l\, .
 \label{t-cond-1D}
 \end{equation}
 Consequently the condition (\ref{A-cond}) is equivalent to the following one
 \begin{equation}
 \frac{\pp{f_l}{u_k}}{f_l-f_k}=\frac{\pp{\lambda_l}{u_k}}{\lambda_l-\lambda_k}\, ,\qquad l \neq k \, .
 \label{A-1Dcond}
 \end{equation}
 Equations (\ref{Rinv1D}) represent the well known form of the one-dimensional multi-component hydrodynamic type systems in terms of Riemann invariants (see e.g.
 \cite{L-VI,Whi}). Hodograph equation (\ref{hodo1D}) and condition (\ref{A-1Dcond}) are exactly those of the Tsarev generalized hodograph method
 \cite{Tsa,DN}.\par 
 
 So, solutions of the homogeneous Euler equations (\ref{g-Hopf-g}), for which functions $f_l$, $l=1,\dots,n$ in (\ref{hodo-g}) are selected according to the condition 
 (\ref{A-1Dcond}), after the restriction to the one-dimensional subspace $x_1=\dots=x_n$ become the solutions of the system (\ref{Rinv1D}).\par
 
 It is noted that reduction of the homogeneous Euler equation to the system (\ref{Rinv1D})    arises also for other one-dimensional restrictions of the 
 $n-$dimensional space $(x_1,\dots,x_n)$, for instance, given by  $x_2=x_3=\dots=x_n=0$. In these cases the characterizations of functions $f_l$ are quite different from 
 (\ref{A-1Dcond}).  
\section{$(n-1)$-dimensional reductions: Jordan system}
\label{sec-Jordan}
Here we will consider reductions of the Euler system ({\ref{hodo-g}}) with $\lambda_k(\uu)=u_k$ to the $(n-1)$-dimensional 
subspace defined by the restriction $x_n=0$. It is equivalent to require $\alpha_n=0$ in the hodograph equations (\ref{hodo-g}). The relation 
(\ref{udx}) implies that $\pp{u_l}{x_n}=0$, $l=1,\dots,n$ under this restriction, however due to (\ref{udx}) 
\begin{equation}
\frac{1}{\alpha_n}\pp{u_l}{x_n}= - (A^{-1})_{ln}\neq 0\, , \qquad l=1,\dots,n\, .
\end{equation}  
Using this relation, one rewrites equation (\ref{g-Hopf-g}) as ($\alpha_k=\beta_k=1$, $k=1,\dots,n-1$, $\beta_n=1$)
\begin{equation}
\begin{split}
\pp{u_l}{t}+\sum_{k=1}^{n-1} u_k \pp{u_l}{x_k} - u_n (A^{-1})_{ln}&=0 \, , \qquad l=1,\dots,n-1 \\
\pp{u_n}{t}+\sum_{k=1}^{n-1} u_k \pp{u_n}{x_k} - u_n (A^{-1})_{nn}&=0\, .
\label{Hopf-n-1-r}
\end{split}
\end{equation}
Under the requirements 
\begin{equation}
(A^{-1})_{ln}=-\pp{u_n}{x_l}\, , \qquad l=1,\dots, n-1
\label{Arl}
\end{equation}
and 
\begin{equation}
(A^{-1})_{nn}=0
\label{Arn}
\end{equation}
the system (\ref{Hopf-n-1-r})  assumes the form 
\begin{equation}
\begin{split}
\pp{u_l}{t}+\sum_{k=1}^{n-1} u_k \pp{u_l}{x_k} + u_n \pp{u_n}{x_l}&=0 \, , \qquad l=1,\dots,n-1 \\
\pp{u_n}{t}+\sum_{k=1}^{n-1} u_k \pp{u_n}{x_k} &=0\, .
\end{split}
\end{equation}
In terms of variables $u_i$, $i=1,\dots, n-1$ and $v=u_n^2/2$, the system looks like
\begin{equation}
\begin{split}
\pp{u_l}{t}+\sum_{k=1}^{n-1} u_k \pp{u_l}{x_k} +  \pp{v}{x_l}&=0 \, , \qquad l=1,\dots,n-1 \\
\pp{v}{t}+\sum_{k=1}^{n-1} u_k \pp{v}{x_k} &=0\, .
\end{split}
\label{Jordan-nD}
\end{equation}
The system (\ref{Jordan-nD}) represents the $(n-1)-$dimensional generalization of the one-dimensional $(n=2)$ Jordan system introduced in \cite{KK}.\par

The system (\ref{Jordan-nD}) at $n=4$ arises also in physics. Indeed, hydrodynamical equations describing  adiabatic flow of an ideal fluid are of the form \cite{Lamb,L-VI}
\begin{equation}
\begin{split}
\pp{u_l}{t}+\sum_{k=1}^{3} u_k \pp{u_l}{x_k} +\frac{1}{\rho}  \pp{P}{x_l}&=0 \, , \qquad l=1,2,3 \\
\pp{S}{t}+\sum_{k=1}^{3} u_k \pp{S}{x_k} &=0\, .
\end{split}
\label{adiab-3D}
\end{equation}
 where $\rho$ is the fluid density, $P$ stands for pressure and $S$ is the entropy. The variation of enthalpy $W$ is given by (see e.g. \cite{L-VI})
 \begin{equation}
 \pp{W}{x_i}=T \pp{S}{x_i}+\frac{1}{\rho} \pp{P}{x_i}\, , \qquad i=1,2,3\, .
 \end{equation} 
 So for the isoenthalpic motion with constant temperature one has
 \begin{equation}
 \frac{1}{\rho}\pp{P}{x_i}=\pp{}{x_i}(TS)\, , \qquad i=1,2,3
 \label{therm-rel}
 \end{equation}
and, consequently, one immediately concludes that the system (\ref{Jordan-nD}) at $n=4$ and $v=-TS$ describes the adiabatic and isoenthalpic motion of a fluid at
constant temperature.\par

Now let us analyze  the conditions (\ref{Arl}) and (\ref{Arn}). The relation (\ref{udx}) says that 
\begin{equation}
\pp{u_n}{x_l} =-(A^{-1})_{nl}\, ,\qquad l=1,\dots,n-1\, ,
\end{equation}
and so the condition (\ref{Arl}) is satisfied if 
 \begin{equation}
 (A^{-1})_{nl}=(A^{-1})_{ln}\, ,\qquad l=1,\dots,n-1\, .
 \label{An-symm}
 \end{equation}
 Thus, equations (\ref{g-Hopf-g}) are reducible to (\ref{Jordan-nD})  if the matrix $A_{lk}=\pp{g_l}{u_k}$ obeys the constraints (\ref{An-symm}), (\ref{Arn}) or,
 equivalently 
 \begin{equation}
 \begin{split}
 \tilde{A}_{ln}&=\tilde{A}_{nl}\, , \qquad l=1,\dots,n-1\\
 \tilde{A}_{nn}&=0\, ,
 \end{split}
 \label{tilde-cond}
 \end{equation}
where $\tilde{A}$ is the matrix adjugate to $A$ (i.e. $A \tilde{A}=\det(A) {I}_n$).\par 

Using the known formula for the adjugate matrix $\tilde{A}$, one can obtain more explicit form of the conditions (\ref{tilde-cond}). 
We instead will use an explicit form of the matrix $A^{-1}$. Indeed, since $A_{lk}=\pp{g_l}{u_k}$, one has
\begin{equation}
(A^{-1})_{lk}=\pp{u_l}{g_k}\ , \qquad l,k=1, \dots n\, .
\label{invA-cond}
\end{equation} 
Using (\ref{invA-cond}), one rewrites (\ref{tilde-cond}) as 
\begin{equation}
\begin{split}
\pp{u_l}{g_n}&=\pp{u_n}{g_l}\, , \qquad l=1,\dots,n-1\\
\pp{u_n}{g_n}&=0\, .
\end{split}
\label{ug-cond}
\end{equation}
Conditions (\ref{ug-cond}) imply that 
\begin{equation}
\begin{split}
u_l=&\pp{}{g_l}\phi(g_1,\dots,g_{n})+\mathcal{A}_l(g_1,\dots,g_{n-1}) \, ,  \qquad l=1,\dots,n-1\, , \\
u_n=&\pp{}{g_n}\phi(g_1,\dots,g_{n}) \, , 
\end{split}
\end{equation}
where $\phi(g_1,\dots,g_{n})$ and $\mathcal{A}_l(g_1,\dots,g_{n-1})$ are arbitrary functions. \par

Consider now $1-$form ($x_i$, $i=1,\dots,n$ and $t$ are fixed)  
\begin{equation}
\Omega=\sum_{l=1}^n u_l \D g_l=\D \phi +\sum_{l=1}^{n-1} \mathcal{A}_l(g_1,\dots,g_{n-1}) \D g_l
\label{Omega-1} 
\end{equation}
and perform the Legendre transformation defined by 
\begin{equation}
\sum_{l=1}^n g_l \D u_l=\D \left( \sum_{l=1}^n  u_l g_l \right)-\Omega\, .
\end{equation}
Due to (\ref{Omega-1}) one has
\begin{equation}
\sum_{l=1}^n g_l \D u_l=\D W - \sum_{l=1}^{n-1} \mathcal{A}_l(g_1,\dots,g_{n-1}) \D g_l\, ,
\label{Legendre}
\end{equation}
where
\begin{equation}
W= \left( \sum_{l=1}^n  u_l g_l \right)- \phi\, .
\end{equation}
Equation (\ref{Legendre}) rewritten as
\begin{equation}
\sum_{l=1}^n \left(g_l -\pp{W}{u_l} +\sum_{k=1}^{n-1} \mathcal{A}_k \pp{g_k}{u_l}\right)\D u_l=0\, ,
\end{equation}
implies that
\begin{equation}
g_l =\pp{W}{u_l} -\sum_{k=1}^{n-1} \mathcal{A}_k \pp{g_k}{u_l}\, , \qquad l=1,\dots,n\, .
\label{pot-gl}
\end{equation}
The compatibility condition for (\ref{pot-gl}) (i.e.  $ \partial^2W/\partial u_l \partial u_k= \partial^2W/\partial u_k \partial u_l$) is given by
\begin{equation}
\pp{g_l}{u_m}-\pp{g_m}{u_l}+\sum_{k,i=1}^{n-1} \left( \pp{\mathcal{A}_k}{g_i}- \pp{\mathcal{A}_i}{g_k}\right)\pp{g_i}{u_m}\pp{g_k}{u_l}=0\, , \qquad l,m=1,\dots,n\, .
\label{comp-gen-J}
\end{equation}
Correspondingly for $f_l$ one has
\begin{equation}
f_l =\pp{\tilde{W}}{u_l} -\sum_{k=1}^{n-1} \tilde{\mathcal{A}}_k \pp{f_k}{u_l}\, , \qquad l=1,\dots,n\, .
\label{pot-fl}
\end{equation}
and
\begin{equation}
\pp{f_l}{u_m}-\pp{f_m}{u_l}+\sum_{k,i=1}^{n-1} \left( \pp{\tilde{\mathcal{A}}_k}{g_i}- \pp{\tilde{\mathcal{A}}_i}{g_k}\right)\pp{f_i}{u_m}\pp{f_k}{u_l}=0\, , \qquad l,m=1,\dots,n\, ,
\label{comp-gen-J-f}
\end{equation}
where $\tilde{W}(u_1,\dots,u_n)$ and $\tilde{\mathcal{A}}_l(f_1,\dots,f_{n-1})$, $l=1,\dots,n-1$ are arbitrary functions.\par

Equations  (\ref{pot-gl}) and (\ref{comp-gen-J}) or (\ref{pot-fl}) and (\ref{comp-gen-J-f}) plus the condition (\ref{Arn})
characterise those solutions of the homogeneous Euler equations which are, at the same time, solutions of the $(n-1)$-dimensional Jordan system (\ref{Jordan-nD}) 
or  the system (\ref{adiab-3D}). It is noted that in this case the class of solutions is parametrized by one arbitrary function $\tilde{W}(u_1,\dots, u_n)$ (or $W$)
on $n$ variables and $(n-1)$ functions $\tilde{\mathcal{A}}_l(f_1,\dots,f_{n-1})$ of $n-1$ variables.
\section{Potential flows}
\label{sec-potJ}
The formulae presented in the previous section are simplified drastically in the particular case when
\begin{equation}
\pp{\mathcal{A}_k}{g_i}=\pp{\mathcal{A}_i}{g_k}\, , \qquad i=1,\dots,n-1\, ,
\end{equation}
or, consequently
\begin{equation}
\mathcal{A}_i=\pp{\psi}{g_i}\, ,\qquad i=1,\dots,n-1\, ,
\end{equation}
where $\psi$ is an arbitrary function. So $g_l=\pp{}{u_l}(W-\psi)$. Equivalently without loose of generalities one can put directly $\mathcal{A}_i \equiv 0$. In this case 
\begin{equation}
g_l=\pp{W}{u_l}\, , \qquad l=1,\dots,n\, ,
\label{gWpot}
\end{equation}
and the matrix $A$ is of the form
\begin{equation}
A_{lk}=\frac{\partial^2 W}{\partial u_l \partial u_k}\, , \qquad l,k=1,\dots,n\, .
\label{A-Hess}
\end{equation}
Then, one has
\begin{equation}
(A^{-1})_{nn}=\frac{\det B}{\det A} \, , 
\label{Ann-BA}
\end{equation}
where $(n-1)\times (n-1)$ matrix $B$ is the algebraic complement to the element  $A_{nn}$, i.e.
\begin{equation}
B_{lk}=\frac{\partial^2 W}{\partial u_l \partial u_k}\, , \qquad l,k=1,\dots,n-1\, .
\label{BW}
\end{equation}
So, the condition (\ref{Arn}) assumes the form
\begin{equation}
\det (B)=0\, .
\label{Bcond}
\end{equation} 
The form (\ref{A-Hess}) of the matrix $A$ leads to a constraint of the variables $u_i$, $i=1,\dots,n$. Indeed, since the matrix $A$   (\ref{A-Hess}) is symmetric one,
then the matrix $A^{-1}$ is symmetric too. In such a case the relations (\ref{udx}) imply that
\begin{equation}
\pp{u_l}{x_k}=\pp{u_k}{x_l}\, , \qquad k,l=1,\dots, n-1\, .
\label{upotc}
\end{equation}
 So 
 \begin{equation}
 u_i=\pp{\phi}{x_i}\, , \qquad i=1,\dots, n-1\, ,
 \label{upot}
 \end{equation}
 where $\phi(x_1,\dots,x_{n-1})$ is some function. Thus in this case the equations (\ref{Jordan-nD}) or  (\ref{adiab-3D}) describe the potential adiabatic 
 isoenthalpic flows.\par 
 
 Due to (\ref{upot}) equations  (\ref{Jordan-nD})  are equivalent to the following (assuming that all constants of integration vanish) 
 \begin{equation}
 \begin{split}
 & \pp{\phi}{t}+\frac{1}{2} \sum_{k=1}^{n-1} \left( \pp{\phi}{x_k} \right)^2 + v=0\, , \\
 & \pp{v}{t}+ \sum_{k=1}^{n-1} \pp{\phi}{x_k}\pp{v}{x_k} =0\, .
 \end{split}
\label{potsys}
 \end{equation}
 The first equation (\ref{potsys}) is well known in hydrodynamics, e.g. for the  isoentropic potential motion (see \cite{L-VI},  \S 5,109). \par
 
 In our case the elimination of $v$ from the system (\ref{potsys}) gives us the following equation for the velocity potential $\phi$
 \begin{equation}
 \frac{\partial^2 \phi}{\partial t^2} + 2 \sum_{k=1}^{n-1} \pp{\phi}{x_k}  \frac{\partial^2 \phi}{\partial x_k \partial t}
 +  \sum_{i,k=1}^{n-1} \pp{\phi}{x_k}\pp{\phi}{x_i}  \frac{\partial^2 \phi}{\partial x_k \partial x_i}=0 \, .
 \label{poteqn}
 \end{equation}
Solutions of this equation provide us the solutions of the system (\ref{potsys}) via $v= -\pp{\phi}{t}-\frac{1}{2} \sum_{k=1}^{n-1} \left( \pp{\phi}{x_k} \right)^2 $. \par

The system (\ref{potsys}) can be viewed also as the Hamilton-Jacobi equation given by the first of equations (\ref{potsys}) for the action $\phi$ 
with the time-dependent potential 
$v(x_1,\dots,x_{n-1};t)$ obeying the second equation (\ref{potsys}). So the solutions of the system (\ref{potsys}) provide us with a solvable $(n-1)-$dimensional system of
classical mechanics.\par

In the one dimensional case $n=2$ the equation (\ref{poteqn}) is of the form
 \begin{equation}
 \frac{\partial^2 \phi}{\partial t^2} + 2  \pp{\phi}{x}  \frac{\partial^2 \phi}{\partial x \partial t}
 +  \left(\pp{\phi}{x}\right)^2  \frac{\partial^2 \phi}{\partial x^2}=0 \, .
 \label{poteqn-1D}
 \end{equation}
  or
 \begin{equation}
 \pp{}{t} 
 \left(\pp{\phi}{t} +  \left(\pp{\phi}{x}\right)^2 \right)
 + \pp{}{x} \left(\frac{1}{3} \left(\pp{\phi}{x}\right)^3  \right)=0 \, .
 \label{poteqn-1D-cons}
 \end{equation}
  and it is of parabolic type.\par
  
  Solutions of the system (\ref{potsys}) and equation (\ref{poteqn})  are provided by hodograph equations (\ref{hodo-g}) with function $g_l$ obeying
  (\ref{gWpot}) and (\ref{BW}), (\ref{Bcond}).\par
  
One can obtain the corresponding expression also for the functions $f_l$, $l=1,\dots,n$. 
Indeed since $g_l=x_l-u_l t +f_l$, the 1-form 
\begin{equation}
\tilde{\Omega}=\sum_{l=1}^n  {f_l} \D u_l=-\Omega^*-\D \left( \sum_{l=1}^n x_l u_l - \frac{t}{2} \sum_{l=1}^n  u_l^2 \right)
\end{equation}
is closed due to the fact that $\Omega^*=\D W$.
Consequently one has
\begin{equation}
f_l=\pp{\tilde{W}}{u_l}\, , \qquad l=1,\dots,n
\end{equation}
 for some function $\tilde{W}(u_1,\dots,u_n)$.  In terms of the function $\tilde{W}$ the condition (\ref{Bcond}) looks like
\begin{equation}
\det (\tilde{B})=0\, .
\label{tBcond}
\end{equation} 
where
\begin{equation}
\tilde{B}_{lk}=\frac{\partial^2 \tilde{W}}{\partial u_l \partial u_k } -t \delta_{lk}\, , \qquad l,k=1, \dots, n-1\, .
\label{Beledefn}
\end{equation}

Thus, for the $(n-1)$-dimensional Jordan system (\ref{Jordan-nD}) and the system (\ref{adiab-3D}) and (\ref{therm-rel}), hodograph equations (\ref{hodo-g}) 
$(\lambda_i=u_i)$  are the equations $\pp{W}{u_i}=0$, $i=1,\dots,n$ defining the critical points of functions $W=W(\uu,\underline{x},t)$ of the form $(v=u_n^2/2)$
\begin{equation}
W=\sum_{i=1}^{n-1} x_i u_i- t\left(v+ \frac{1}{2} \sum_{i=1}^{n-1} u_i^2 \right)+\tilde{W}(\uu,v)
\label{Wgen-J}
\end{equation}
with functions $\tilde{W}$ obeying the constraint (\ref{tBcond}) at the critical points. The potential Jordan system (\ref{Jordan-nD}) describes the dynamics of the critical points
of such functions $W$. \par

In other words, solutions of the potential Jordan system (\ref{Jordan-nD}) or equations (\ref{adiab-3D}) and (\ref{therm-rel}) are those solutions of the homogeneous Euler 
equation  (\ref{g-Hopf-g}) which corresponds to the functions $f_i$ in (\ref{hodo-g}) being the components of the gradients of functions $W$ of the form (\ref{Wgen-J})
 with obeying  the constraints (\ref{tBcond}). \par
 
 The condition that $\det(\tilde{B})$ belongs to the ideal generated by the functions $\pp{W}{u_i}$, $i=1,\dots,n$ is the sufficient one to characterize the above 
 subclasses of solutions. However, more explicit description of the class of the class of the functions $W$ would be, definitely, rather convenient.\par
 
 For this purpose we first observe that at the critical point $t=\pp{\tilde{W}}{v}$. So the matrix $\tilde{B}$ can be equivalently rewritten as
 \begin{equation}
 \tilde{B}_{lk}=\frac{\partial^2 \tilde{W}}{\partial u_l \partial u_k}-\delta_{lk} \pp{\tilde{W}}{v}\, ,\qquad l,k=1,\dots,n-1\, .
 \end{equation}
Then since
 \begin{equation}
\frac{\partial^2 \tilde{W}}{\partial u_l \partial u_k}-\delta_{lk} \pp{\tilde{W}}{v}=
\frac{\partial^2 W}{\partial u_l \partial u_k}-\delta_{lk} \pp{W}{v}\, ,
 \end{equation}
the condition (\ref{tBcond}) is equivalent to
\begin{equation}
\det \left( \frac{\partial^2 W}{\partial u_l \partial u_k}-\delta_{lk} \pp{W}{v} \right)=0\, .
\label{WcondJ}
\end{equation}
The last step is to extend this condition outside the critical points and to consider (\ref{WcondJ}) as the equation defining the function $W$ of the form
(\ref{Wgen-J}). For such functions conditions  (\ref{tBcond}) and (\ref{Beledefn}) are automatically satisfied at the critical points. Note that the formula (\ref{Wgen-J})
and (\ref{WcondJ}) are natural $(n-1)-$dimensional extensions of the corresponding formulae $\pp{W}{v}=\frac{\partial^2 W}{\partial u_1^2}$ 
for the one-dimensional Jordan system \cite{KK}. \par

We note also that one gets the same system (\ref{Jordan-nD}) considering the restrictions to the $(n-1)-$dimensional subspaces defined by conditions $x_i=x_k$
with fixed $i$ and $k$. In these cases one has characterizations of the functions $f_i$ similar to those considered above. \par

Finally we note that the subclass of the hodograph equations (\ref{hodo-g}) with $g_i = \pp{W}{u_1}$, $i=1,\dots,n$ and, hence, $\lambda_i = \pp{F}{u_i}$, $i=1,...,n$ where 
$W$ and $F$ are some functions give us solutions of the potential reduction $u_i = \pp{\phi}{x_1}$, $i=1,\dots,n$ of the homogeneous Euler equation (\ref{g-Hopf-g}). 
In this case equation (\ref{g-Hopf-g}) is equivalent to the following ($\alpha_i = \beta_i =1$, $i=1,\dots,n$)
\begin{equation}
  \pp{\phi}{t} +   F \left( \pp{\phi}{x_1}, \dots, \pp{\phi}{x_n}\right) = 0 \, .
\end{equation}

\section{$(n-1)$-dimensional reductions: Polytropic gas}
\label{sec-Poly}
Now we will consider equations (\ref{Hopf-n-1-r}) with the constaints
\begin{equation}
(A^{-1})_{ln} =-u_n^a \pp{u_n}{x_l}\, , \qquad l=1,\dots,n-1\, ,
\label{polyA-c}
\end{equation}
and
\begin{equation}
(A^{-1})_{nn} =-\sum_{k=1}^{n-1}  \pp{u_k}{x_k}\, ,
\label{polyA-cn}
\end{equation}
where $a$ is an arbitrary real number. Under these constraints equations (\ref{Hopf-n-1-r}) assume the form
\begin{equation}
\begin{split}
&\pp{u_l}{t}+\sum_{k=1}^{n-1}u_k \pp{u_l}{x_k}=-\frac{1}{a+2} \pp{}{x_l} \left( u_n^{a+2} \right)\, , \qquad l=1,\dots,n-1\, , \\
& \pp{u_n}{t}+\sum_{k=1}^{n-1}\pp{}{x_k}(u_n u_k)=0\, .
\end{split}
\label{polyeqn}
\end{equation}
For $a=-1$ this system represents the shallow water equation in $(n-1)-$dimension with $u_n$ being the fluid height $h$. For arbitrary $a$ it describes the polytropic 
motion with pressure $p=\frac{1}{a+3}\rho^{a+3}$ and the density $\rho=u_n$ (see e.g. \cite{L-VI,Whi}).\par

The formula (\ref{udx}) implies that
\begin{equation}
\pp{u_n}{x_l}=-(A^{-1})_{nl}\, , \qquad l=1,\dots,n-1\, ,
\end{equation}
and
\begin{equation}
\sum_{k=1}^{n-1} \pp{u_k}{x_k}=-\sum_{k=1}^{n-1}(A^{-1})_{kk}\, .
\end{equation}
So, the constraint (\ref{polyA-c}) and (\ref{polyA-cn}) are equivalent to the following 
\begin{equation}
\begin{split}
(A^{-1})_{ln}&=u^a_n(A^{-1})_{nl} \, , \qquad l=1,\dots,n-1\\
(A^{-1})_{nn}&=\sum_{k=1}^{n-1}(A^{-1})_{kk}\, .
\end{split}
\label{poly-constr}
\end{equation}
Using (\ref{invA-cond}), one rewrites these constraints as
\begin{equation}
\begin{split}
\pp{u_l}{g_n}&=u^a_n\pp{u_n}{g_l}\equiv\pp{}{g_l}\frac{u_n^{a+1}}{a+1}\, , \qquad l=1,\dots,n-1\\
\pp{u_n}{g_n}&=\sum_{k=1}^{n-1}\pp{u_k}{g_k}\, .
\end{split}
\label{upolycon}
\end{equation}
 The first condition (\ref{upolycon}) imply that
 \begin{equation}
 \begin{split}
 u_l&=\pp{\phi}{g_l}+\mathcal{B}_l(g_1,\dots,g_{n-1})\, , \qquad l=1,\dots,n-1\, , \\
 \frac{1}{a+1} u_n^{a+1}&=\pp{\phi}{g_n}\, ,
 \end{split}
 \label{u-potgen}
 \end{equation}
where $\phi(g_1,\dots,g_{n})$  and  $\mathcal{B}_l(g_1,\dots,g_{n-1})$ are arbitrary functions. On can find the corresponding formulae and constraints for $g_l$ in a 
way similar to that described in section \ref{sec-Jordan}. Here we will consider the simplest case $\mathcal{B}_l(g_1,\dots,g_{n-1})=0$, $ l=1,\dots,n-1$. In this case 
one has
\begin{equation}
 \begin{split}
 g_l&=\pp{W}{u_l}\, , \qquad l=1,\dots,n-1\, , \\
 g_n&=u_n^{-a}\pp{W}{u_n}\, ,
 \end{split}
 \label{u-potgen-B0}
\end{equation}
for some function $W$. \par

So the matrix $A$ is of the form
\begin{equation}
A=
\left(
\begin{array}{ccc}
B&&V_1 \\
&& \\
V_2 && \pp{}{u_n} \left( u_n^{-a} \pp{W}{u_n}\right)
\end{array}
\right)\, .
\label{Apolyred}
\end{equation}
where $B$ is a $(n-1) \times (n-1)$ matrix with elements $\frac{\partial^2 W}{\partial u_l \partial u_k} $, 
$V_1$ is a column with  $(n-1)$ elements $ \frac{\partial^2 W}{\partial u_l \partial u_n}$,
$V_2$ is a row with  $(n-1)$ elements $u_n^{-a }\frac{\partial^2 W}{\partial u_n \partial u_l}$.
Hence, the second condition (\ref{poly-constr}) assumes the form 
\begin{equation}
\det B - \sum_{k=1}^{n-1} \det C_k=0\, ,
\label{W-poly-cond}
\end{equation}
where $C_k$ are algebraic complements of the elements $A_{kk}$. \par

So, the solutions of the system (\ref{polyeqn}) describing polytropic motion are those solutions of the homogeneous $n-$dimensional Euler equation 
which correspond to the choice (\ref{u-potgen-B0}) of functions $g_l$ with $W$ obeying the constraint (\ref{W-poly-cond})  
on the manifold $g_i=0$, $i=1,\dots  n$.  \par

Analogously to the previous section one can show that the functions $f_l$ are given by 
\begin{equation}
 \begin{split}
 f_l&=\pp{\tilde{W}}{u_l}\, , \qquad l=1,\dots,n-1\, , \\
 f_n&=u_n^{-a}\pp{\tilde{W}}{u_n}\, ,
 \end{split}
\end{equation}
for some function $\tilde{W}$. So the function $W$ is of the form 
\begin{equation}
W=\sum_{i=1}^{n-1} x_i u_i -t\left( \frac{1}{2} \sum_{k=1}^{n-1} u_k^2 +\frac{1}{a+2} u_n^{a+2}  \right) +\tilde{W}\, .
\end{equation}
Since
\begin{equation}
\frac{\partial^2 W}{\partial u_k \partial u_l}=-t \delta_{lk}+\frac{\partial^2 \tilde{W}}{\partial u_k \partial u_l}\, , \qquad l,k=1\dots,n-1
\end{equation}
and 
\begin{equation}
t=u_n^{-1-a} \pp{\tilde{W}}{u_n}\, , 
\end{equation}
the condition (\ref{W-poly-cond}) is equivalent to
\begin{equation}
\det \left( \frac{\partial^2 \tilde{W}}{\partial u_l \partial u_k}-  u_n^{-1-a} \delta_{lk} \pp{\tilde{W}}{u_n}\right) -\sum_{k=1}^{n-1} \det \tilde{C}_k=0\, ,
\label{tW-cond}
\end{equation}
where $ \tilde{C}_k$ are principal minors of the matrix
\begin{equation}
\tilde{A}_{lk}= \frac{\partial^2 \tilde{W}}{\partial u_l \partial u_k}-  u_n^{-1-a} \delta_{lk} \pp{\tilde{W}}{u_n}\, .
\label{Atmp}
\end{equation}
Finally using the relation
\begin{equation}
 \frac{\partial^2 {W}}{\partial u_l \partial u_k}-  u_n^{-1-a} \delta_{lk} \pp{{W}}{u_n}=
  \frac{\partial^2 \tilde{W}}{\partial u_l \partial u_k}-  u_n^{-1-a} \delta_{lk} \pp{\tilde{W}}{u_n}\, , \qquad l,k=1,\dots,n-1\, ,
\end{equation}
one can extend the equation (\ref{tW-cond}) outside the critical points $\pp{W}{u_i}=0$ to obtain the equation characterizing the function $W$, i.e.
\begin{equation}
\det \left( \frac{\partial^2  {W}}{\partial u_l \partial u_k}-  u_n^{-1-a} \delta_{lk} \pp{{W}}{u_n}\right) -\sum_{k=1}^{n-1} \det {C}_k=0\, ,
\label{W-cond-noncrit}
\end{equation}
where $C_k$ are principal minors of the matrix (\ref{Atmp}) with the substitution $\tilde{W}\to W$.\par

In the simplest case $n=2$ all above formulae become rather compact. For arbitrary $a$ the function $W$ is of the form $(u_1=u,\, u_2=v)$.\par

\begin{equation}
W=xu-t\left(\frac{1}{2}u^2+\frac{1}{a+2} v^{a+2}\right)+\tilde{W}
\end{equation}
while the equation (\ref{W-cond-noncrit}) becomes
\begin{equation}
\frac{\partial^2 W}{\partial u^2}-v^{-a} \frac{\partial^2 W}{\partial v^2}+a v^{-1-a} \pp{W}{v}=0\, .
\label{W1D}
\end{equation}
For $a=-1$ this equation is quite similar to that associated with  is isentropic motion of fluid (see e.g. \cite{L-VI}, \S 105).
In the one dimensional case the system (\ref{polyeqn}) is diagonalizable to the following
\begin{equation}
\pp{\Gamma_\pm}{t}=\lambda_{\pm} \pp{\Gamma_{\pm}}{x}
\end{equation}
  with  the Riemann invariants $\Gamma_\pm$ and the characteristics velocities $\lambda_\pm$ given by \cite{Whi}
  \begin{equation}
  \Gamma_\pm= -\frac{a+2}{2}u\pm v^{\frac{a+2}{2}}\, , \qquad \lambda_\pm= -u\pm v^{\frac{a+2}{2}}\, .
  \end{equation}
In terms of the Riemann invariants, the equation (\ref{W1D}) becomes the classical Euler-Poisson-Darboux equation
\begin{equation}
(\Gamma_+-\Gamma_-) \frac{\partial^2 W}{\partial u \partial v} =-\frac{a}{a+2} \left(\pp{W}{\Gamma_+}-\pp{W}{\Gamma_-}\right)\, .
\label{W-EPD}
\end{equation} 
For the classical shallow water equation $a=-1$ equation (\ref{W-EPD}) coincides with that studied in \cite{KMAM}.\par

We see that equations characterizing functions $W$ are nonlinear for multi-dimensional systems (\ref{Jordan-nD}) and (\ref{polyeqn}), in contrast to the 
one dimensional situation with the linear equation $\pp{W}{v}=\frac{\partial^2 W}{ \partial u_1^2}$ and equation (\ref{W-EPD}). Such situation seems to be typical in 
applications of the hodograph equation to multi-dimensional PDEs \cite{MS1,MS2}, except, of course, the master homogeneous Euler equation (\ref{g-Hopf-g}).\par

We note also that one can study in a similar manner the dimensional reductions of the generalised equation (\ref{g-Hopf-g}) with arbitrary 
function $\lambda_i(\underline{u})$.
\section{Reductions of the three-dimensional Euler equations}
\label{sec-3Ered}
In this section we consider two particular examples of the Euler equation  in  three dimensions. \par

First, let us start with the two-dimensional restriction of the hodograph equation given by $(x_1=x_2=x,\, x_3=y)$,
\begin{equation}
\begin{split}
x-\lambda_1 t +f_1=0\, , \\
x-\lambda_2 t +f_2=0 \, , \\
y-\lambda_3 t +f_3=0 \, .
\end{split}
\label{hodo-2s}
\end{equation}
Differentiation of (\ref{hodo-2s}) w.r.t. $x,y$ and $t$ gives
\begin{equation}
\begin{split}
\pp{u_l}{x}&=-(A^{-1})_{l1}-(A^{-1})_{l2} \, ,\\
\pp{u_l}{y}&=-(A^{-1})_{l3}\, , \\
\pp{u_l}{t}&=\sum_{k=1}^3 (A^{-1})_{lk} \lambda_k \, , \qquad l=1,2,3 \, .
\end{split}
\label{A-3-eqn}
\end{equation}
Combining expressions (\ref{A-3-eqn}), one obtains 
\begin{equation}
\begin{split}
\pp{u_1}{t} +\lambda_1 \pp{u_1}{x}+\lambda_3 \pp{u_1}{y}&=(\lambda_2-\lambda_1)(A^{-1})_{12}\, , \\
\pp{u_2}{t} +\lambda_2 \pp{u_2}{x}+\lambda_3 \pp{u_2}{y}&=(\lambda_1-\lambda_2)(A^{-1})_{21}\, , \\
\pp{u_3}{t} +\lambda_3 \pp{u_3}{x}+\lambda_1 \pp{u_3}{y}&=(\lambda_2-\lambda_1)(A^{-1})_{32}\, .
\end{split}
\label{A-3-evol}
\end{equation}
 Imposing the constraint
 \begin{equation}
 (A^{-1})_{12}=(A^{-1})_{21}=(A^{-1})_{32}=0\, ,
 \label{A-3-c}
 \end{equation}
one gets the system
\begin{equation}
\begin{split}
\pp{u_1}{t} +\lambda_1 \pp{u_1}{x}+\lambda_3 \pp{u_1}{y}&=0\, , \\
\pp{u_2}{t} +\lambda_2 \pp{u_2}{x}+\lambda_3 \pp{u_2}{y}&=0\, , \\
\pp{u_3}{t} +\lambda_3 \pp{u_3}{x}+\lambda_1 \pp{u_3}{y}&=0\, ,
\end{split}
\label{es-3-2D}
\end{equation}
which is two dimensional extension of the one-dimensional system for Riemann invariant $u_1$ and $u_2$.\par

Constraints (\ref{A-3-c}) are equivalent to the following three equations for three  functions $g_1,g_2,g_3$
\begin{equation}
\begin{split}
\pp{g_1}{u_3}\pp{g_3}{u_2}-\pp{g_1}{u_2}\pp{g_3}{u_3}&=0 \\
\pp{g_2}{u_3}\pp{g_3}{u_1}-\pp{g_2}{u_1}\pp{g_3}{u_3}&=0 \\
\pp{g_1}{u_2}\pp{g_3}{u_1}-\pp{g_1}{u_1}\pp{g_3}{u_2}&=0 
\end{split}
\label{g-3-con}
\end{equation}
In terms of the functions $f_i$ one has the equations (\ref{g-3-con}) with the substitution ($t=(f_1-f_2)/(\lambda_1-\lambda_2)$)
\begin{equation}
\pp{g_l}{u_k}=\pp{f_l}{u_k}-\frac{f_1-f_2}{\lambda_1-\lambda_2} \pp{\lambda_l}{u_k}\, , \qquad l,k=1,2,3\, .
\end{equation}
So any solution of the three-dimensional homogeneous Euler equation, constructed using the functions $f_i$, $i=1,2,3$ obeying equations (\ref{g-3-con}),
is a solution of the two-dimensional system (\ref{es-3-2D}). \par

We note that the system (\ref{es-3-2D})   does not reduce to the expression (\ref{Rinv1D}) in the naive one dimensional limit $x=y$. The reason is that
the constraint (\ref{A-3-c}) represent only the part of the constraint (\ref{A-cond}).\par

In order to recover this system we combine the expressions (\ref{A-3-eqn}) into another system  of equations (equivalent to (\ref{A-3-c})), namely,
 \begin{equation}
\begin{split}
\pp{u_1}{t} +\lambda_1 \pp{u_1}{x}&=(\lambda_2-\lambda_1)(A^{-1})_{12} + \lambda_3 (A^{-1})_{13}  \, , \\
\pp{u_2}{t} +\lambda_2 \pp{u_2}{x}&=(\lambda_1-\lambda_2)(A^{-1})_{21} + \lambda_3 (A^{-1})_{23} \, , \\
\pp{u_3}{t} +\lambda_3 \pp{u_3}{x}&=\lambda_1 (A^{-1})_{31} + \lambda_2 (A^{-1})_{32}  \, . 
\end{split}
\label{A-3-1D}
\end{equation}
Now, requiring that $x=y$ and $(A^{-1})_{lk}=0$, $l \neq k$, $l,k=1,2,3$, i.e. $A_{lk}=0$, $l \neq k$, one obtains the 3-component system (\ref{Rinv1D}). \par

As second example we consider the one-dimensional reduction of the Euler equation with $\lambda_k=u_k$ and $\alpha_1=1$, $\alpha_2=\alpha_3=0$  and 
$\beta_1=\beta_2=1$, $\beta_3=0$. So we start with the hodograph system
\begin{equation}
\begin{split}
x-u_1 t+f_1&=0\, , \\
- t+f_2&=0\, , \\
f_3&=0 \, ,
\end{split}
\label{hodo-2exe}
\end{equation}
where $x=x_1$, $x_2=x_3=0$ and we redefine the function $f_2\to u_2 f_2$. Differentiating (\ref{hodo-2exe}) w.r.t. $x$ and $t$, we obtain
\begin{equation}
\begin{split}
\pp{u_l}{x}&=-(A^{-1})_{l1}\, , \qquad l=1,2,3\, , \\
\pp{u_l}{t}&=(A^{-1})_{l1}u_1+(A^{-1})_{l2}     \, .
\end{split}
\label{uA-2exe}
\end{equation}
Consequently one has the system
\begin{equation}
\pp{u_l}{t}+u_1\pp{u_l}{x}=(A^{-1})_{l2} \, , \qquad l=1,2,3\, .
\end{equation}
Now we impose the constraints 
\begin{equation}
(A^{-1})_{12} =-\pp{u_2}{x}\, , \qquad (A^{-1})_{22}=-\pp{u_3}{x}\, , \qquad (A^{-1})_{32}=0\, .    
\end{equation}
Due to the relation (\ref{uA-2exe}), these constraints are equivalent to the following
\begin{equation}
(A^{-1})_{12} = (A^{-1})_{21}\, , \qquad (A^{-1})_{22}=(A^{-1})_{31}\, , \qquad (A^{-1})_{32}=0\, .    
\end{equation}
Using the explicit form of the $3\times 3$ inverse of the matrix $A$, one obtains the following system of equations
\begin{equation}
\begin{split}
\pp{g_1}{u_3}\pp{g_3}{u_2}-\pp{g_2}{u_3} \pp{g_3}{u_1} +\pp{g_3}{u_3} \left(\pp{g_2}{u_1}-\pp{g_1}{u_2}\right) &=0\\
\pp{g_1}{u_1}\pp{g_3}{u_3}-\pp{g_2}{u_1} \pp{g_3}{u_2} +\pp{g_3}{u_1} \left(\pp{g_2}{u_2}-\pp{g_1}{u_3}\right) &=0\\
\pp{g_1}{u_2}\pp{g_3}{u_1}-\pp{g_1}{u_1} \pp{g_3}{u_2} &=0\, .
\end{split}
\label{exe2-g}
\end{equation}
 Any solution of this system with the substitution 
 \begin{equation}
 \pp{g_1}{u_l}= \pp{f_1}{u_l}-f_2 \delta_{1l}\, , \qquad 
 \pp{g_2}{u_l}= \pp{f_2}{u_l}\, , \qquad  \pp{g_3}{u_l}= \pp{f_3}{u_l}\, , 
 \end{equation}
provide us with the functions $f_1,f_2,f_3$ for which three dimensional homogeneous Euler equations is reducible to the system
\begin{equation}
\begin{split}
\pp{u_1}{t}+u_1\pp{u_1}{x}+\pp{u_2}{x}&=0 \, , \\
\pp{u_2}{t}+u_1\pp{u_2}{x}+\pp{u_3}{x}&=0 \, , \\
\pp{u_3}{t}+u_1\pp{u_3}{x}&=0 \, , 
\end{split}
\end{equation}
which is the $3$-component one-dimensional Jordan system described in \cite{KK}. \par

It is not difficult to show that the system (\ref{exe2-g}) has a solution for which 
\begin{equation}
g_i=\pp{W}{u_i}\, ,\qquad i=1,2,3
\end{equation}
where the function $W$ obeys the equation
\begin{equation}
\pp{W}{u_2}=\frac{\partial^2 W}{\partial u_1^2}\, , \qquad 
\pp{W}{u_3}=\frac{\partial^3 W}{\partial u_1^3}\, .
\label{W23}
\end{equation}
Hodograph equations (\ref{hodo-2exe}) represent the critical points equations $\pp{W}{u_i}=0$, $i=1,2,3$ for the function
\begin{equation}
W=x u_1-t \left(\frac{1}{2}u_1^2+u_2\right)+\tilde{W}(u_1,u_2,u_3)\, ,
\label{W-1}
\end{equation}
which obeys the equations (\ref{W23}).

Equations (\ref{W23}) and function $W$ (\ref{W-1}) are exactly those given in the paper \cite{KK}.
\section{Infinite-dimensional Euler equation: reductions to Jordan chain, Burgers and Korteweg-de Vries equations}
\label{sec-infEred}
The above result on the $3$-component Jordan system can be extended to the $n$-component case. Indeed, let us consider the hodograph equations for arbitrary $n$ and
$\alpha_1=1$,  $\alpha_2=\dots=\alpha_n=0$,  $\beta_1=\beta_2=1$, $\beta_3=\dots=\beta_n=0$, i.e. the equations ($x=x_1$)
\begin{equation}
\begin{split}
x-u_1t+f_1=&0\, ,\\
-t+f_2=&0\, ,\\
f_m=&0\, , \qquad m=3,\dots,n 
\end{split}
\end{equation}
where, for convenience we redefine the function $f_2 \to u_2 f_2$. Relation (\ref{udx}) imply that $\pp{u_l}{x_k}=0$, $k=2,3,\dots,n$ and
\begin{equation}
\begin{split}
\pp{u_l}{x}&=-(A^{-1})_{l1}\, , \\
\pp{u_l}{t}&=(A^{-1})_{l1} u_1 +(A^{-1})_{l2}\, . 
\end{split}
\label{u-A-nc}
\end{equation}
Combining (\ref{u-A-nc}), one gets 
 \begin{equation}
 \pp{u_l}{t}+u_1 \pp{u_l}{x}= (A^{-1})_{l2}\, , \qquad l=1,\dots,n\, .
 \end{equation}
Imposing the constraint 
\begin{equation}
\begin{split}
(A^{-1})_{l2}&=-\pp{u_{l+1}}{x}\, , \qquad l=1,\dots,n-1\, , \\ 
(A^{-1})_{n2}&=0\, ,
\end{split}
\label{A-r-chain}
\end{equation}
one obtains the $n$-components system
\begin{equation}
\begin{split}
 \pp{u_l}{t}+u_1 \pp{u_l}{x}+\pp{u_{l+1}}{x}&= 0\, , \qquad l=1,\dots,n-1\, , \\
 \pp{u_n}{t}+u_1 \pp{u_n}{x}&= 0\, ,
\end{split}
\label{Jcn}
\end{equation}
that is the $n$-component Jordan system introduced in \cite{KK}.\par

Since $\pp{u_{l+1}}{x}=-(A^{-1})_{l+1,1}$, $l=1,\dots,n-1$, the constraints (\ref{A-r-chain}) are equivalent to the following 
\begin{equation}
\begin{split}
(A^{-1})_{l2}&=(A^{-1})_{l+1,1}\, , \qquad l=1,\dots,n-1\, , \\ 
(A^{-1})_{n2}&=0\, ,
\end{split}
\end{equation}
or 
\begin{equation}
\begin{split}
\tilde{A}_{l2}&=\tilde{A}_{l+1,1}\, , \qquad l=1,\dots,n-1\, , \\ 
\tilde{A}_{n2}&=0\, .
\end{split}
\label{tA-cchain}
\end{equation}
Using the explicit expression for the elements of the adjugate matrix $\tilde{A}$, one rewrites the constraints (\ref{tA-cchain}) as the system of $n$ differential  equations for the functions $g_i$, $i=1,\dots, n$ or $f_i$, $i=1,\dots, n$. It is not difficult to show that this system has a solution for which
\begin{equation}
g_i=\pp{W}{u_i}\, , \qquad f_i=\pp{\tilde{W}}{u_i}\, , \qquad i=1,\dots,n
\end{equation}
where the functions $W$ and $\tilde {W}$ obey the equations
\begin{equation}
\pp{W}{u_k}=\frac{\partial^k W}{ \partial u_1^k}\, , \qquad k=2,\dots,n\, ,
\end{equation}
and
\begin{equation}
W=x u_1 -t \left( \frac{1}{2} u_1^2+u_2\right) +\tilde{W}(u_1,\dots, u_n)\, ,
\end{equation}
that coincides with those formulae presented in \cite{KK}. It is noted that in this one-dimensional reduction the constraint (\ref{upotc})  is absent. \par

Now, following \cite{KK} one can consider the system (\ref{Jcn}) in the formal limit $n \to \infty$ and get the infinite Jordan chain which has been discussed in different contexts
in \cite{KMAM,Pav,KK}.  So the Jordan chain represent a particular reduction of the infinite-dimensional Homogeneous Euler equation.\par

In the paper \cite{KO} it was observed that the Jordan chain admits   differential reductions to various integrable partial differential equations, for example, to the Burgers 
equation and Korteweg-de Vries equation. Indeed, if one imposes the constraint 
\begin{equation}
u_2=\pp{u_1}{x}\, ,
\label{conBur}
\end{equation}
then the first equation $(l=1)$ in (\ref{Jcn}) becomes the Burgers equation
\begin{equation}
\pp{u_1}{t}+u_1 \pp{u_1}{x}+\frac{\partial^2 u_1}{\partial u_1^2}=0\, ,
\end{equation}
while the other equations (\ref{Jcn}) with $l=2,3,\dots$ represent themselves the recursive relations to define $u_3,u_4,\dots$. \par 

If one requires that 
\begin{equation}
u_2=\frac{\partial^2 u_1}{\partial x^2}\, ,
\label{conKdV}
\end{equation}
then the Jordan chain is reduced to the Korteweg-de Vries equation
\begin{equation}
\pp{u_1}{t}+u_1 \pp{u_1}{x}+\frac{\partial^3 u_1}{\partial u_1^3}=0\, .
\end{equation}

Constraints (\ref{conBur}) and (\ref{conKdV}) can be rewritten in terms of the elements $(A^{-1})_{lk}$. Indeed, the differential consequence of (\ref{conBur}),
 namely,  $\pp{u_2}{x}=\pp{}{x} \left(\pp{u_1}{x}\right)$ after the use  of (\ref{udx}), assumes the form
 \begin{equation}
 (A^{-1})_{21}+\sum_{k=1}^\infty \pp{(A^{-1})_{11}}{u_k} (A^{-1})_{k1}=0\, .
 \label{conABur}
 \end{equation}
The differential consequence of (\ref{conKdV}) is equivalent to the following 
\begin{equation}
(A^{-1})_{21} - \sum_{k,l=1}^{\infty} \pp{}{u_l} \left(\pp{(A^{-1})_{kl}}{u_k} (A^{-1})_{k1}\right) (A^{-1})_{l1}=0\, .
 \label{conAKdV}
\end{equation}
Though the constraint (\ref{conABur}), (\ref{conAKdV}) are rather cumbersome, one concludes that the solutions of the Burgers and Korteweg-de Vries 
equations represent particular subclasses of solutions of the infinite-dimensional homogeneous Euler equation.
\section{Gradient catastrophe for the homogeneous Euler equations}
\label{sec-catE}
All the results presented in the previous sections are valid under the assumption that $\det(A) \neq 0$. If instead
\begin{equation}
\det(A) \equiv \det \left( \pp{f_l}{u_k}-t \pp{\lambda_l}{u_k}\right)=0\, ,
\label{gcat}
\end{equation}
then, according to (\ref{udx}), (\ref{udt}) solutions of the equation (\ref{g-Hopf-g}) and other equations exhibit the gradient catastrophe  $\pp{u_l}{x_k}\to \infty$,
$\pp{u_l}{t}\to \infty$. \par

Let us consider such a situation for the classical homogeneous Euler equation ($\lambda_k=u_k$, $\alpha_k=\beta_k=1$). 
In this case the equation (\ref{gcat}) is simplified to
\begin{equation}
\det(A) \equiv \det \left( \pp{f_l}{u_k}-t \delta_{lk}\right)=0\, ,
\label{Ecat}
\end{equation}
i.e. to the characteristic polynomial equation
\begin{equation}
t^n+\sum_{k=0}^{n-1} B_k(\underline{u}) t^k=0
\label{tc-pol}
\end{equation}
of the $n \times n$ matrix $\tilde{A}_{lk}=\pp{f_l}{u_k}$. 
Due to (\ref{0hodo}) the functions $\underline{f}(\underline{u})$ are the local inverse  of the initial values of $\underline{u}(t=0,\underline{x})$  and, 
  consequently, coefficients $B_k$ depends on $u_1,\dots,u_n$ only. \par
  
 Thus, gradient catastrophe for the homogeneous Euler equation happens in general on the $n-$dimensional hypersurface in $\mathbb{R}^{n+1}$  given by the
 equation (\ref{tc-pol}). If the polynomial (\ref{tc-pol}) has no real roots then the gradient catastrophe does not happen for given initial data $u_i(t=0,x)$. Let us assume that
 equation (\ref{tc-pol}) has at least one real root $t_c$. So the gradient catastrophe happens on the hypersurface $\mathcal{S}$ given by
 \begin{equation}
 t_c=\phi(u_1,\dots,u_n)
 \label{tc-surf}
 \end{equation}
where $\phi(\underline{u})$ is a certain function constructed out from the (local) inverse of the $\underline{u}(t=0,\underline{x})$. Usually discussed first moment 
of appeareance of gradient catastrophe corresponds to the minimum value of $t_c$, i.e. to the situation when
\begin{equation}
\pp{t_c}{u_i}=\pp{}{u_i}\phi(u_1,\dots,u_n)=0\, , \qquad i=1,\dots,n\
\label{t-crit}
\end{equation}
plus a condition on the second derivatives  (for generic catastrophes such condition is given by the classical condition on the Hessian of $\phi$).\par

For generic initial data the function $\phi$ is generic. Consequently $n$ equations (\ref{t-crit}) has generically a single solution $u_1^c,\dots,u_n^c$.\par

Thus, generically, the gradient catastrophe for the homogeneous Euler equation first happens at the time
 \begin{equation}
 {t_c}_\mathrm{min}=\phi(u_1^c,\dots,u_n^c)
 \label{tcat}
 \end{equation}
 at the point $u_1^c,\dots, u_n^c$ on the hypersurface $\mathcal{S}$ (\ref{tc-surf}). Then it expands on the whole hypersurface (\ref{tc-surf}). \par
 
 It is noted that for the first time such property of the gradient catastrophe for multi-dimensional equations has been observed in \cite{MS1,MS2}.\par
 
In more detail the gradient catastrophes for the homogeneous Euler equation,  related  equations and its regularization will be considered in a separate paper.
\subsubsection*{Acknowledgement} The authors are thankful to M. Pavlov for the information about D. B. Fairlie's papers. This project thanks the support of the European Union's Horizon 2020 research and innovation programme under the Marie Sk{\l}odowska-Curie grant no 778010 {\em IPaDEGAN}. We also gratefully acknowledge the auspices of the GNFM Section of INdAM under which part of this work was carried out. 

\end{document}